\documentclass[galaxies,communication,accept,moreauthors,pdftex]{Definitions/mdpi} 
\usepackage{xcolor}
\firstpage{1} 
\pubvolume{1}
\issuenum{1}
\articlenumber{0}
\pubyear{2023}
\copyrightyear{2023}
\externaleditor{Academic Editor: Firstname Lastname} 
\datereceived{19 October 2023} 
\daterevised{16 November 2023} 
\dateaccepted{18 November 2023} 
\datepublished{ } 
\hreflink{https://doi.org/} 
\Title{On the Maximum Energy Release from Formation of Static Compact Objects}

\TitleCitation{On the Maximum Energy Release from Formation of Static Compact Objects}

\Author{Abhas Mitra 
$^{1,}$*\orcidA{}, Krishna Kumar Singh $^{1,2,}$*\orcidB{}}
\AuthorNames{Abhas Mitra and Krishna Kumar Singh}

\AuthorCitation{Mitra, A.; Singh, K.K.}

\address{%
$^{1}$ \quad Homi Bhabha National Institute, Anushakti Nagar, Mumbai 400094, India\\
$^{2}$ \quad Astrophysical Sciences Division, Bhabha Atomic Research Centre, Mumbai 400085, India}
\corres{Correspondence: akmitra@hbni.ac.in (A.M.);~kksastro@barc.gov.in (K.K.S.)}
\abstract
{\textls[-15]{Type II Supernova 1987A (SN 1987A), observed in 1987, released an energy of  \mbox{$Q \approx 3 \times 10^{53}$ erg}}.
This huge energy is essentially the magnitude of gravitational potential  or self-gravitational energy (PE) 
of a new born cold  neutron star having a gravitational compactness or redshift $z_b \approx 0.15$. One may wonder what could 
be the upper limit on the amount of energy that might be released with the formation of  a cold Ultra Compact Object 
(UCO) with an arbitrary high $z_b$. Accordingly, here, for the first time, we obtain an analytical expression for the 
PE of a homogeneous general relativistic UCO assuming it to be cold and static. It is found that the PE of a homogeneous UCO 
of mass $M$ may exceed Mc$^2$ and be as large as 1.34 Mc$^2$. This result, though surprising, follows from an 
\textit{exact and correct} analytical calculation based on the standard General Theory of Relativity (GTR).  Further,
UCOs supported by tangential stresses may be inhomogeneous and much more massive than neutron stars with PE $\sim$ 2.1 Mc$^2$ Thus, in principle, formation of  
an UCO of a few solar masses ($M_\odot$) might release an energy  $Q\sim10^{55}$ erg.} 


\begin{document}
\section{Introduction}\label{sec1}
Compact objects are mostly formed through the gravitational collapse of normal stars at the end of their lives. White Dwarfs, Neutron Stars, 
and Black Holes (BHs) belong to the family of relativistic compact objects in astrophysics. Static or non-rotating white dwarfs, with~a radius 
comparable to that of Earth, have masses lower than the well-known Chandrashekhar mass-limit of 1.4 $M_\odot$ \cite{Chandrasekhar1931}. 
Neutron stars, with~a typical radius of 10 km, have masses up to the Oppenheimer-Volkoff limit of 2--3 $M_\odot$ \cite{Oppenheimer1939}. 
Above this limit, stars may collapse into black holes, whose radius is given by the radius of event horizon $R_g = 2M$, in~natural 
units with $G = c = 1$. Studies on radiative spherical general relativistic collapse suggest that the fluid becomes hotter during the 
collapse~\cite{Herrera2006}. Two equations of states of the perfect fluid, based on the formulations using baryonic number density 
and mass density, are used in the literature to study the binding energy of static fluids~\cite{Karkowski2004}. The~latter approach is 
preferred in the problems of General Relativity (GR). Polytropic equations of state are also used to calculate the potential energy, ($\Omega$), 
of the fluids~\cite{Karkowski2004}. Emission of electromagnetic photons, neutrinos, and~gravitational waves is believed to take place during 
the general relativistic collapse and the radiative energy increases steadily~\cite{Herrera2006,Mitra2006}. 
\par
The enormous amounts of energy, observed in various astrophysical energetic phenomena are believed to be powered by the compact objects 
from the interactions with strong and large-scale magnetic fields on their surrounding magnetospheres. More than three decades ago, the~
titanic Supernova 1987A (SN 1987A) was discovered on 23 February, 1987 in the nearby Large Magellanic Cloud, which is a satellite galaxy of 
the Milky-Way at a distance of 160 kilo-light years. This was the first time a burst of cosmic neutrinos ($\nu$), accompanying the stellar 
collapse, was detected on Earth from beyond our solar system~\cite{Hirata1987}. Even after more than three decades, this remains the only case 
to support the idea that general relativistic gravitational collapses are accompanied by the emission of electromagnetic radiation and neutrinos. 
Multi-messenger observations including gravitational waves are expected to provide more insights into the modern supernova models~\cite{Damiano2023}. 
Long duration Gamma Ray Bursts (GRBs) are the most energetic explosions in the Universe involving isotropic equivalent gamma ray energy release of 
10$^{51-55}$ erg over a duration of $T_{90} > 2s$. In~general, more than 15 such GRBs have been observed so far~\cite{Burns2023}. It is widely believed 
that long duration GRBs are produced in the collapse of a massive star and this may lead to the black holes or neutron stars/magnetars 
having a physical surface. The~gravitational or magnetic energy release in these cosmic explosions is speculated to power the bright $\gamma$-ray emission 
as well as the cosmic ray acceleration~\cite{Waxman1995}. They are considered as a possible source of high energy neutrinos. Neutrino bursts could 
be the primary source of energy emission in other astrophysical scenarios as well~\cite{Belczynski2008,Zenati2023}.
\par
SN 1987A also confirmed the prediction on the occurrence of neutrino-antineutrino ($\nu -\bar \nu$) bursts by the theories of supernova involving 
stellar core collapse~\cite{Burrows1986,Woosley1986,Arnett1989}. The~measured energy of the $\nu$-burst, (2.9 $\pm$ 1.2) $\times 10^{53}$ erg, can be 
understood as the \textit{magnitude} of the potential energy ($Q = |\Omega|$), as~the positive energy is liberated to compensate for 
the increase of the (negative) self-gravitational energy. By~using the equation given below, such an equality ($Q = |\Omega|$) suggests  that the 
gravitational mass of the new born neutron star to be (1.38 $\pm$ 0.43)M$_\odot$ where M$_\odot$ is the solar mass:
\begin{equation}
	Q= |\Omega| \approx  \frac{\rm GM^2}{R}~=~ 3 \times 10^{53} \left({M\over 1.4~M_\odot}\right)^2 \left({10~{\rm km} \over R}\right)~{\rm erg}
\end{equation}

For proper appreciation of the foregoing equation, we need to briefly revisit the idea of polytropes, self-gravitating fluid spheres. The~equation 
of state connecting the (isotropic) pressure (P) and density ($\rho$) satisfies
\begin{equation}
	P = K \rho^\gamma~~,
\end{equation}
where $K$ is a constant. If~the fluid sphere is an inhomogeneous polytrope of index $n$, then $\gamma = (1 +1/n)$ \cite{Chandrasekhar1939}. 
It turns out that the case of a homogeneous sphere corresponds to $n=0, \gamma=\infty$. For~a polytrope having index $n$, it is found that
\begin{equation}\label{poly-ene}
	|\Omega| = {3 \over 5-n} {GM^2 \over R}
\end{equation}

Clearly, the~uniform density case of $|\Omega| = (3/5) GM^2/R$ corresponds to $n=0$ case. The~internal energy density ($u$) of the polytropic fluid 
is~\cite{Chandrasekhar1939}
\begin{equation}
u = {P \over \gamma -1}
\end{equation}

Then it is seen that, for~uniform density case with $\gamma =\infty$, $u=0$ and so is the thermal internal energy $U=0$ despite arbitrary high pressure.
A related concept here is the Newtonian total energy of the {system}
 ($E^{N}$) which is the sum of the negative self-gravitational energy 
($\Omega$) and positive internal/thermal energy ($U$). $E^N$ is given as~\cite{Weinberg1972}
\begin{equation}
E^N = U~+~ \Omega
\end{equation}

For a Newtonian polytrope, one finds  that~\cite{Weinberg1972}
\begin{equation}
	E^{N} = {n-3 \over 5-n} {GM^2\over R} 
\end{equation}

When $u=U=0$, one has $E^N = \Omega$, and~from energy conservation, the~energy liberated in the formation of the compact object
\begin{equation}
Q = |E^N| = |\Omega| = {3 \over 5} {GM^2 \over R}
\end{equation}

The key idea here is that for the sake of the principle of energy conservation, an~amount of energy equal to the negative $\Omega$ must be radiated 
away in some form or other. This point will become clearer when we will move to a general relativistic viewpoint of this issue.
However,  here we run into a small problem because the neutron star mass in SN 1987A was determined from Equation~(1) : $\Omega = GM^2/R$. 
The only  way to resolve this contradiction is that the cold compact object is not homogeneous, but~has a structure similar to a $n =3/2$~polytrope.
\begin{equation}
	|\Omega| = {3\over 3.5} {GM^2 \over R}  \approx {GM^2 \over R} 
\end{equation}
And we shall make this revision for the case of General Relativity too. Now let us dig deeper into this idea.
\par
By comparing  Equations~(7) and (8), one can find that
\begin{equation}
 	|\Omega|^N (Observation) = 1.6 |\Omega|^N (Hom)
\end{equation}

Essentially, $|\Omega|$ increases significantly with extreme inhomogeneity because more and more mass lie in the deeper potential well as compared to 
the homogeneous case. And~this effect of enhancement in PE with inhomogeneity is expected to be more pronounced for general relativistic case for the extreme 
non-linearity of General Relativity.
\par
Having made this discussion in the framework of simple Newtonian gravitation, we shall now explore the question of highest possible value of $|\Omega|$ 
by using the GTR. Accordingly, for~the first time, we shall work out an exact expression for general relativistic $|\Omega|$ for a homogeneous fluid 
sphere in Section~\ref{sec2}.
\section{Potential Energy in General~Relativity}\label{sec2}
While in Newtonian physics, $\rho$ represents the baryonic mass alone, in~GTR, $\rho$ includes contribution from internal energy as well. 
It is well known that, the~gravitational or ADM (Arnowitt, Deser and Misner) mass of the fluid sphere is given as~\cite{Tooper1964,Weinberg1972,Misner1973,Shapiro1983}
\begin{equation}
	M = \int_0^R \rho ~4\pi   r^2  ~ dr 
\end{equation}
where $r$ is the areal radius or Schwarzschild radial coordinate for a static case. Unlike the Newtonian case, in~GTR case, $\rho$ includes all 
internal and radiation energies along with the baryonic energy density. Also, the~total mass energy is $E=Mc^2$ in contrast to the Newtonian total 
energy $E^N$. However $M$  is not the sum of the locally added mass-energies because  the proper radial distance $dl \neq dr$:
\begin{equation}
	dl = \sqrt{-g_{rr}} ~ dr  =   {dr  \over \sqrt{1 - 2m/r}}~~,
\end{equation}
where
\begin{equation}
	m(r) = \int_0^r \rho ~4 \pi   r^2 dr~~,
\end{equation}
is the gravitational mass of an interior spherical section. Accordingly,  the~proper mass or the sum of locally added mass-energies in the curved 
space-time  is~\cite{Tooper1964,Weinberg1972,Misner1973,Shapiro1983}
\begin{equation}
	M_p = \int_0^R \rho ~4\pi r^2  \sqrt{-g_{rr}} ~ dr = \int_0^R {4\pi \rho  r^2  \over \sqrt{1 - 2m/r}} ~dr
\end{equation}

Since for the  asymptotically flat space-time,  $\sqrt{- g_{rr}} \to 1$ as $r \to \infty$, proper mass is the initial or original gravitational mass 
at $r\to \infty$. In~GTR, the~fundamental definition of PE is given by (see Equation~(3).31 in~\cite{Tooper1964}):
\begin{equation}
\Omega = M~-~M_p
\end{equation}

Clearly, $M < M_p$ because self-gravitational energy is negative. And~this shows the energy released in the process of gravitational collapse. We shall 
assume here that the  compact object is  born in a catastrophic collapse and the new born object is static and cold, i.e.,~its internal energy is negligible 
compared to its proper mass energy $U \ll M_p$ (this is the case for white dwarfs and neutron stars). In~such a case the energy released in the collapse 
process will be:
\begin{equation}
	Q~=~M_p - M = |\Omega| = \int_0^R 4\pi  \rho r^2 { dr \over \sqrt {1- 2m/r}} - {M}
\end{equation}

In the Newtonian limit of $2m/r \ll 1$, one finds $\sqrt{-g_{rr}} \approx (1+ m/r)$, so that
\begin{equation}
	Q=~|\Omega|~= \int_0^M   {m dm\over r}
\end{equation}
which for $\rho = constant$ case reduces to $|\Omega| = (3/5) M^2/R$. 
\par
Equation~(15) cannot be integrated for an inhomogeneous sphere having arbitrary form of  $\rho(r)$. However, for~the first time, we point out that, 
even for a general relativistic case, it is possible to obtain an exact analytical expression for $M_p$ and hence $|\Omega|$ for an uniform density case.
In such a case, we have
\begin{equation}
	M_p =  \int_0^R {4\pi \rho  r^2  \over \sqrt{1 - (8\pi/3)  \rho r^2}} ~dr
\end{equation}

Writing $x = 2m/r = (8\pi/3)  \rho r^2$ and $X = 2M/R = (8\pi/3)  \rho R^2$, one can integrate the foregoing equation:
\begin{equation}
	M_p= {3\over 4} \sqrt { 3\over 8\pi \rho} \left[ \sin^{-1} \sqrt{X} - \sqrt{X} \sqrt {1-X}\right]~~, 
\end{equation}
which may be rewritten as
\begin{equation}
	M_p = {3M\over  2 X^{3/2}} \left[ \sin^{-1} \sqrt{X} - \sqrt{X} \sqrt {1-X}\right] 
\end{equation}

Thus we obtain the \textit{maiden analytical expression} for the PE of a homogeneous fluid sphere in general relativistic case as:
\begin{equation}
	|\Omega| =  {3M\over  2 X^{3/2}} \left[\sin^{-1} \sqrt{X} - \sqrt{X} \sqrt {1-X}\right] - M  = y M
\end{equation}
where
\begin{equation}
	y = {|\Omega| \over M} =   {3\over  2 X^{3/2}} \left[\sin^{-1} \sqrt{X} - \sqrt{X} \sqrt {1-X}\right] - 1
\end{equation}

For sufficiently small values of $X=2M/R$ one can expand
\begin{equation}
	\sin^{-1} \sqrt{X} = \sqrt{X}  +  {X^{3/2} \over 6}  + { 3X^{5/2} \over 40}
\end{equation}
and
\begin{equation}
	\sqrt{X} \sqrt{1-X} = \sqrt{X} - {X^{3/2} \over 2} - {X^{5/2}  \over 8}
\end{equation}

Then one recovers
\begin{equation}
	y = {3\over 10} X = {3\over 5} {M\over R}
\end{equation}
and
\begin{equation}
	|\Omega| = y M = {3\over 5} {M^2 \over R}
\end{equation}

This shows the correctness of Equation~(20) derived above.
\section{Binding Energy of Ultra-Compact~Objects}\label{sec3}
The real measure of the gravitational potential well of ultra-compact objects may be gauged from their surface gravitational redshifts
\begin{equation}
	z_b = (1- 2M/R)^{-1/2} -1 = (1-X)^{-1/2} -1
\end{equation}

Note, for~a typical neutron star, $z_b \approx 0.15$.
\par
For a compact object supported by isotropic pressure alone, the~Buchdahl upper limit of $X = 2M/R= 8/9$ or  $z_b =2.0$ \cite{Buchdahl1959}. 
But for an anisotropic compact object supported partially or fully by tangential pressure too, this upper limit on $X$ increases, and~for 
the case of extreme pressure anisotropy, this \emph{upper limit} is $X \to 1.0$, i.e.,~$z_b \to \infty$\cite{Bowers1974,Lambert2002,Guilherme2019}. 
\par
It may be borne in mind that an exotic compact object (ECO) having $X\to 1$ or $R \to 2M$ is still a \textit{non-singular} object filled with 
matter and having a physical boundary. This is in contrast to the limiting case of $X=1$ or $R=2M$, when one would obtain a singular, vacuum BH 
except for the central singularity.
\par
In fact, it has been claimed that such ECO could be as massive as BHs~\cite{Guilherme2019}. For~a homogeneous ECO, for~the extreme case of  
$X=2M/R \to 1$, one finds that
\begin{equation}
	|\Omega| (Hom) =  \left({3\pi \over 4} -1\right)M \approx 1.34M
\end{equation}

\begin{figure}[H]
\includegraphics[width=0.81\textwidth]{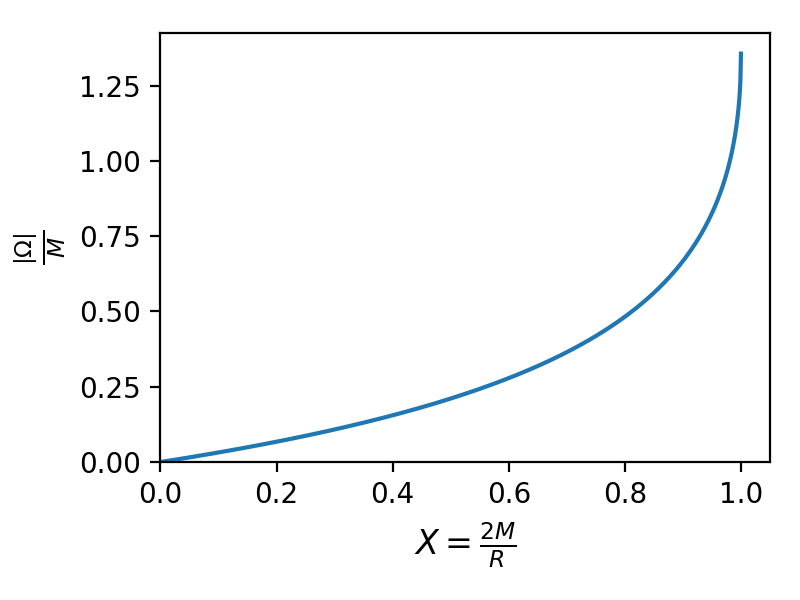}
	\caption{Gravitational Potential Energy Per Unit Mass ($|\Omega|/M$) as a function of Compactness ($X=2M/R$) from Equation~(20).} 
\label{fig}
\end{figure}

In Figure~\ref{fig}, we plot the values of $|\Omega|/M$ against $X=2M/R$ from Equation~(20).

\vspace{-4pt}
If we take the new born neutron star in SN1987A to be of uniform density, then we would obtain a value of $\Omega = 0.6M$.  
For $M =1.4 M$, we would get a value of \mbox{$Q \approx 1.6 \times 10^{53}$ erg}. Thus the assumption of uniform density contradicts 
with the observation that $Q(SN1987A) \approx 3 \times 10^{53}$ erg. For~the GR case too we expect a similar, if~not higher, value 
of $|\Omega|$ compared to the homogeneous case. This is so because relativistic polytropes  have stronger mass concentration towards 
their centers compared to the corresponding Newtonian cases because general relativistic case is much more non-linear than Newtonian gravity. 
Then for the GR UCO too, we may  tentatively write
\begin{equation}
	|\Omega| (Observation) =  1.6 ~ |\Omega| (Hom) \approx 2.1M
\end{equation}

This implies that, in~the extreme case, a~cold  ECO of mass 1$M_\odot$ and extreme pressure anisotropy ($X \to 1$) might be born from an original 
core of mass $M_{p}^{i} \approx 3.1~M_\odot$ with an attendant release of Q = 2.1~M$_\odot c^2$ of~energy.

\par 
While for an extreme pressure anisotropy, it is possible to have $z_b \to \infty$ \cite{Bowers1974,Lambert2002,Guilherme2019}, a~realistic upper limit could 
be $z_b \approx 5.0$ \cite{Ivanov2002,Boehmer2006,Herrera2013}. Table \ref{tab} gives the values of $|\Omega|$ for (i) $z_b = 0.15$, the~neutron star case, (ii)  
$z_b = 2.0$, the~Buchdahl upper limit for isotropic pressure case,  (iii) $z_b = 5.0$, the~realistic upper limit for anisotropic pressure case,~\cite{Ivanov2002,Boehmer2006,Herrera2013}, and~ (iv) $z_b = \infty$, the~theoretical upper limit of  compactness for extremely anisotropic pressure 
case~\cite{Bowers1974,Lambert2002,Guilherme2019}.
\begin{table}[H]
\caption {Magnitude of Gravitational Potential Energy of Highly Inhomogeneous Ultra-Compact Objects.}	\label{tab}
\begin{tabularx}{\textwidth}{CCC}
\toprule
\textbf{z}\boldmath{$_b$}  &\textbf{X = 2M/R} & \boldmath{$|\Omega| (Inhom) $} \\
\midrule
0.15 & $\approx$ 0.15 & 0.15 M\\

2.0  & $\approx$ 0.89 &  1.25 M\\

5.0 & $\approx$ 0.97 &  2.0 M\\

$\infty$ &  1.0 &  2.1 M\\
\bottomrule
\end{tabularx}
\end{table}
\unskip
\section{Discussion and~Outlook} 
General Relativistic astrophysics is almost hundred years old field of research. Even though numerical studies of polytropes 
in General Relativity were first carried out by Tooper long ago~\cite{Tooper1964}, here for the first time, we have obtained 
an exact expression for the gravitational potential energy ($\Omega$) of a homogeneous ultra-compact object. Our exact result 
shows that a homogeneous UCO has an {\em upper limit} of  $|\Omega| (Hom) \approx$ 1.34 Mc$^2$. Here it must be borne in mind 
that though it is possible that  $|\Omega| > M$,  \textit{we always have $|\Omega| < M_p$, the~proper mass-energy of the fluid sphere}.
For highly inhomogeneous case, the~value of $|\Omega|$ increases as more and more mass resides in deep gravitational potential 
well~\cite{Tooper1964}. Observation of SN 1987A suggests that in Newtonian case, $|\Omega^N| (Observation) = 1.6 |\Omega^N| (Hom)$.
For the general relativistic case, this enhancement factor is expected to be larger. Yet if we adopt the enhancement factor of 1.6, 
the highest value of $|\Omega|  (Observation) \approx 2.1Mc^2$.  Indeed an in-depth study of GR gravitational collapse shows that 
the energy emission in the GR case should be higher than the corresponding Newtonian case~\cite{Mitra2006}.

\par
It is important here to mention that neutron stars may be better represented by a polytrope of $n\approx 1.0$ instead of $n=1.5$, 
appropriate for mildly reletivistic degenerate fluid. In this case, the~enhancement factor will become $1.6 \times (3.5/4) \approx 1.3$. 
However, as~before,  because~of extreme nonlinearity of GR, this enhancement factor is expected to be larger than $1.3$. Yet, if~we adopt 
the lower value of $1.3$,  we will have
\begin{equation}
	|\Omega| (n=1.0) =  1.3 ~ |\Omega| (Hom) \approx 1.8 M
\end{equation}
and corresponding $Q= 1.8 M c^2$

\par
Here we may offer yet  another clarification. The~exact upper limit $X=1; ~R=2M$ corresponds to the Schwarzschild black holes which are  vacuum 
except for their central singularity. However, the~limit $X \to 1$  and yet $X <1.0$, may correspond to anisotropic pressure supported 
{\em non-singular} compact objects~\cite{Bowers1974,Lambert2002,Guilherme2019} filled with matter and possessing physical boundaries. 
Technically, a~singular black hole (point singularity) is homogeneous . But~this does not imply that anisotropic non-singular ECOs are 
necessarily homogeneous. On~the other hand, such configurations could to be highly inhomogeneous~\cite{Bowers1974,Lambert2002,Guilherme2019}, 
and~hence the inhomogeneity related modest boosting factors of 1.6 or 1.3 adopted here is justified.
\par
Further, such anisotropic pressure supported compact objects have mass upper limits higher than that of the neutron stars~\cite{Bowers1974,Lambert2002}, and~it has 
even been claimed that they could be as massive as black holes~\cite{Guilherme2019}. Even if we consider the Buchdahl upper limit of compactness for compact 
objects supported by isotropic pressure alone ($z_b = 2.0$), it is possible that $|\Omega| (Inhom) \approx$ 1.25 Mc$^2$.  While for an extreme pressure 
anisotropy, it is possible to have $z_b \to \infty$ \cite{Bowers1974,Lambert2002,Guilherme2019}, the~realistic upper limit could be $z_b \approx 5.0$ 
\cite{Ivanov2002,Boehmer2006,Herrera2013}. After accepting such a realistic upper limit on compactness, it is seen that the birth of a $M=2.5 -3.0 M_\odot$ 
ultra-compact object might be accompanied by the emission of an energy $Q \sim 10^{55}$ erg. This novel result has relevance for most powerful cosmic explosions. 
In~fact the isotropic energy release in the brightest Gamma Ray Burst  GRB 2210009A is $Q\sim 1.5 \times 10^{55}$ erg~\cite{Burns2023}. This enormous isotropic 
energy and close proximity (redshift $\sim$ 0.1505) of the source push the limits of modern theories of GRBs and event rate. Multi-wavelength observations suggest that 
structured jets launched by a common central engine may power the most extreme explosions like GRB 2210009A~\cite{OConnor2023}. One of the important highlights 
of GRB 221009A is the first detection of very high energy $\gamma$-rays above an energy of 10 TeV from a GRB~\cite{Zhen2023}. This provides a unique opportunity 
to explore for possible detection of high-energy neutrinos from GRBs. However, no statistically significant neutrino emission in the energy range 
10$^6$~eV to 10$^{15}$ eV has been reported so far from GRB 221009A during or after the high energy $\gamma$-ray emission~\cite{Abbasi2023}. Only upper limits 
on the time-averaged integral flux of neutrinos are estimated for GRB 221009A. Taking into account the contribution of TeV emission as well as the blast wave 
kinetic energy involved in the afterglow emission, the~isotropic equivalent energy budget of GRB 2210009A corresponds to an energy release of more than 
10$^{55}$ erg~\cite{OConnor2023} which is 30 times higher than the energy released in SN1987A. We remind the reader that we only explored theoretical 
\textit{upper limits} of $Q$ under most favourable cases, and~actual value of $Q$ will be lower in real life (unless $M$ will be higher). 

\par
Finally, whether $Q=1.34 M$ or $Q=2.1 M$ or $Q=1.8 M$, it is always possible that $Q~\approx~10^{55}$ erg because~ECOs supported by extreme anisotropic 
pressure might be arbitrarily massive unlike neutron stars~\cite{Guilherme2019}. Thus, in~principle, it is 
possible to understand the energy budget of GRB 2210009A, $Q \sim 1.5 \times 10^{55}$ erg~\cite{Burns2023}. Accordingly, it seems that, in~future, we might  
hopefully detect cosmic $\nu -\bar \nu$ burst having luminosity  much higher than that of SN 1987A.

\vspace{6pt}
\authorcontributions{conceptualization and  writing--original draft preparation, A.M.; Validation, writing--review and editing, K.K.S. All authors have read and agreed to the published version of the manuscript.}

\funding{This research received no external~funding.}

\dataavailability{No new data were created or analyzed in this study. Data sharing is not applicable to this article.}

\acknowledgments{We thank the anonymous referees for making several constructive suggestions which led to the improvement of the scientific content 
and finally accepting this manuscript for publication.} 

\conflictsofinterest{The authors declare no conflict of~interest.} 


\reftitle{References}

\begin{adjustwidth}{-\extralength}{0cm}

\PublishersNote{}
\end{adjustwidth}



\end{document}